\documentclass[10pt, conference, a4paper]{IEEEtran}

\usepackage{cite}

\ifCLASSINFOpdf
  \usepackage[pdftex]{graphicx}
\else
  \usepackage[dvips]{graphicx}
\fi
\usepackage[tight,footnotesize]{subfigure}
\usepackage{url}

\usepackage[colorinlistoftodos, obeyDraft]{todonotes}

\usepackage{amsmath}

\usepackage{etoolbox}
\makeatletter
\patchcmd{\@makecaption}
  {\scshape}
  {}
  {}
  {}
\makeatother

\def\tablename{Table}

\begin{document}
%
\title{OPS - An Opportunistic Networking Protocol Simulator for OMNeT++ }

\author{\IEEEauthorblockN{Asanga Udugama, Anna F{\"o}rster, Jens Dede, Vishnupriya Kuppusamy and Anas Bin Muslim \vspace{0.2cm}}
\IEEEauthorblockA{Sustainable Communication Networks, University of Bremen, Germany\\
Email: \{adu $|$ anna.foerster $|$ jd $|$ vp $|$ anas1 \}@comnets.uni-bremen.de}
}


%


\maketitle

\begin{abstract}

The number of computing devices of the Internet of Things (IoT) is expected to grow by billions. New networking architectures are being considered to handle communications in the IoT. One of these architectures is Opportunistic Networking (OppNets). To evaluate the performance of OppNets, an OMNeT++ based modular simulator is built with models that handle the operations of the different protocol layers of an OppNets based node. The work presented here provides the details of this simulator, called the Opportunistic Protocol Simulator (OPS).

\end{abstract}


%
\IEEEpeerreviewmaketitle


\section{Introduction}
\label{sec:intro}

The Internet of Things (IoT) is expected to grow into a network of more than 50 billion devices by 2020 \cite{Cisco:2011}. A communication architecture currently being considered for the IoT is Opportunistic Networks (OppNets) \cite{AF:2015}. OppNets are a very versatile and effective means of communication to exchange data in a peer-to-peer manner. There are many application areas in the IoT that benefit from using OppNets. Areas such as social networking, emergencies are where information produced by user devices have a higher value locally than at other locations. OppsNets make it possible to propagate information locally to the interested parties through the different data dissemination protocols and the peer-to-peer link technologies used in most smart devices.

To evaluate the performance of OppNets, we have developed an OMNeT++ based simulator with a number of models. These models implement the functionality of the different layers of an OppNets based node. The purpose of this work is to describe the models available in this simulator, called the Opportunistic Protocol Simulator (OPS) to simulate OppNets.

The rest of this work is ordered in the following manner. The next section (Section~\ref{sec:oppnets}) provides a brief overview to OppNets that also includes an example use case. Section~\ref{sec:related} provides a brief discussion on related work of OMNeT++ based OppNets implementations. Section~\ref{sec:ops} provides the details of the models developed in OPS including the evaluation metrics available in OPS. Section~\ref{sec:eval} provides the details of some evaluations done using OPS. Section~\ref{sec:conclusion} is a concluding summary with a look at future work.


\section{Opportunistic Networks}
\label{sec:oppnets}

The concept of Opportunistic Networking (OppNets) relates to peer-to-peer distribution of information in networksprimarily without the help of any networking infrastructure \cite{Pelusi:2006}. Nodes which are deployed with protocols and mechanisms of OppNets communicate directly with each other when they come into contact. 

OppNets are a very versatile and effective form of networking to be used in a number of application areas ranging from communications during disasters (e.g., South Asian Tsunami) to social networking. They operate on any peer-to-peer communication technology such as Bluetooth and IEEE 802.15.4. A key component of a node in OppNets is the forwarding (i.e., information dissemination) protocols used to propagate information in networks. There are a number of such protocols developed by researchers to efficiently disseminate information throughout a network (e.g., Epidemic Routing \cite{Vahdat:2002}).

\begin{figure}[!ht]
  \centering
    \includegraphics[width=0.45\textwidth]{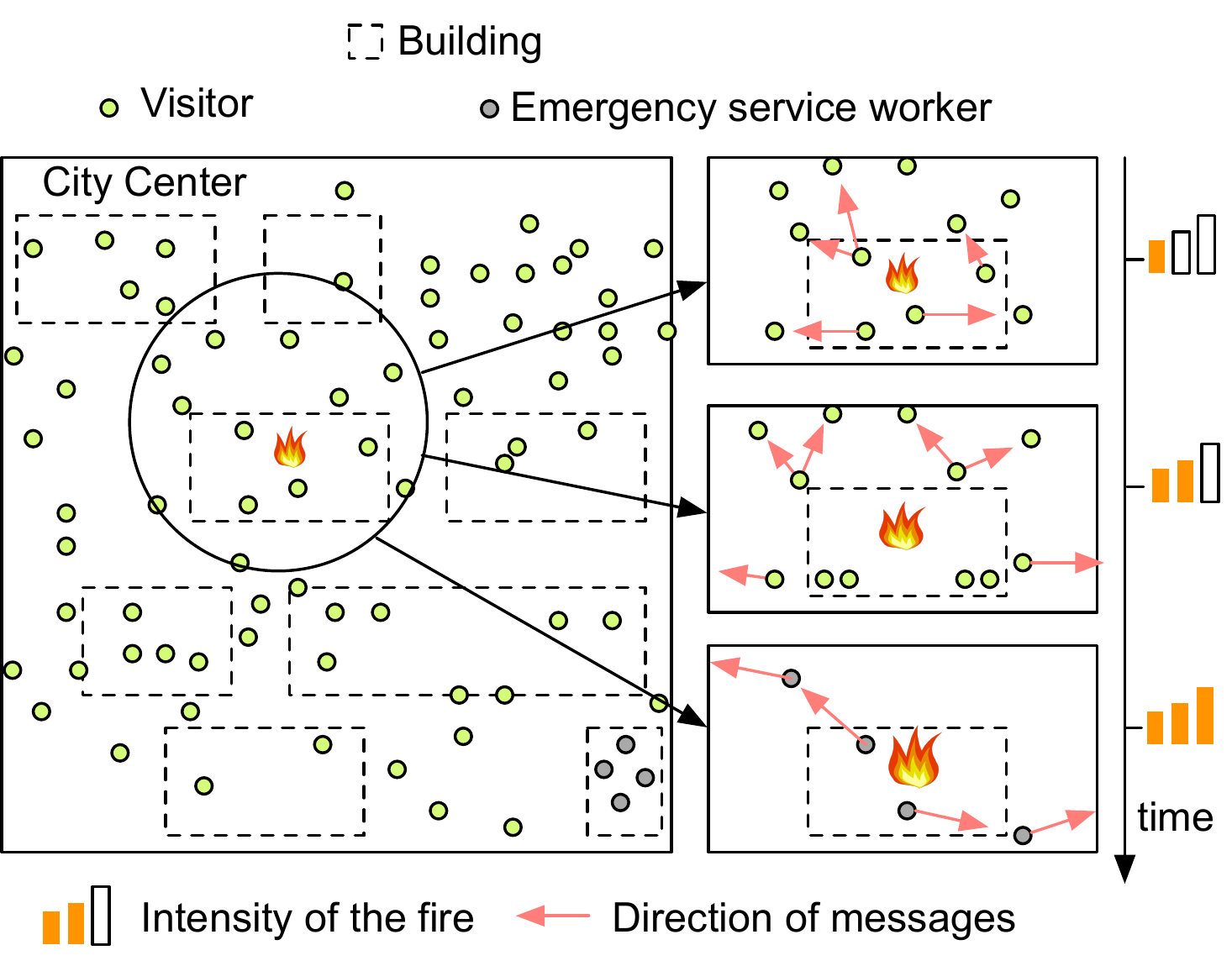}
  \caption{An Example Use Case of OppNets: Dissemination of Information Related to an Emergency}
  \label{fig:oppnets_example}
\end{figure}

Figure~\ref{fig:oppnets_example} shows an example use case of OppNets. A fire breaks out at a shop in the middle of the city center. The visitors and shoppers who are present at that location see this emergency and quickly moves away. The smart devices of these people propagates information related to this emergency over OppNets. This information reaches the emergency service workers who are then able to respond to the emergency.


\section{Related Work}
\label{sec:related}

The OMNet++ simulator and its languages (NED and C++) provide the building blocks to build the layers of functionality of nodes in networks that operate using different wired and wireless protocols. Such protocol implementations are usually made available in OMNeT++ as frameworks. The INET framework is one such framework that provides the implementations for the protocols associated with the Internet Protocol (IP) suite. It is referenced by the official distribution of OMNeT++ and is recommended to be used when simulating IP based networks. 

Though the INET framework consists of many protocols, it lacks the support for protocols required to simulate OppNets. A number of research efforts have focused on developing extensions to the INET framework and other publicly available frameworks to enable OppNets simulations \cite{Helgason:2008, Helgason:2011, Kouyoumdjieva:2012, Zhang:2014, Zhao:2012}. Most of these works have concentrated on improvements to specific areas of OppNets (e.g., mobility, link layer, etc.) and therefore, has resulted in implementations that provide fine grained functionality in those specific areas, leaving out other areas. But, perhaps, the most important two drawbacks of these implementations are - firstly, these extensions have stayed unofficial and secondly, have become obsolete due to the architectural changes in OMNeT++ and the INET framework. Hence the necessity for developing a set of extensible models to build protocols for simulating OppNets in OMNeT++.


\section{Opportunistic Protocol Simulator (OPS)}
\label{sec:ops}

The Opportunistic Protocol Simulator (OPS) is a set of simulation models in OMNeT++ to simulate opportunistic networks. It has a modular architecture where different protocols relevant to opportunistic networks can be developed and plugged in, to evaluate their performance. 

\begin{figure}[!ht]
  \centering
    \includegraphics[width=0.3\textwidth]{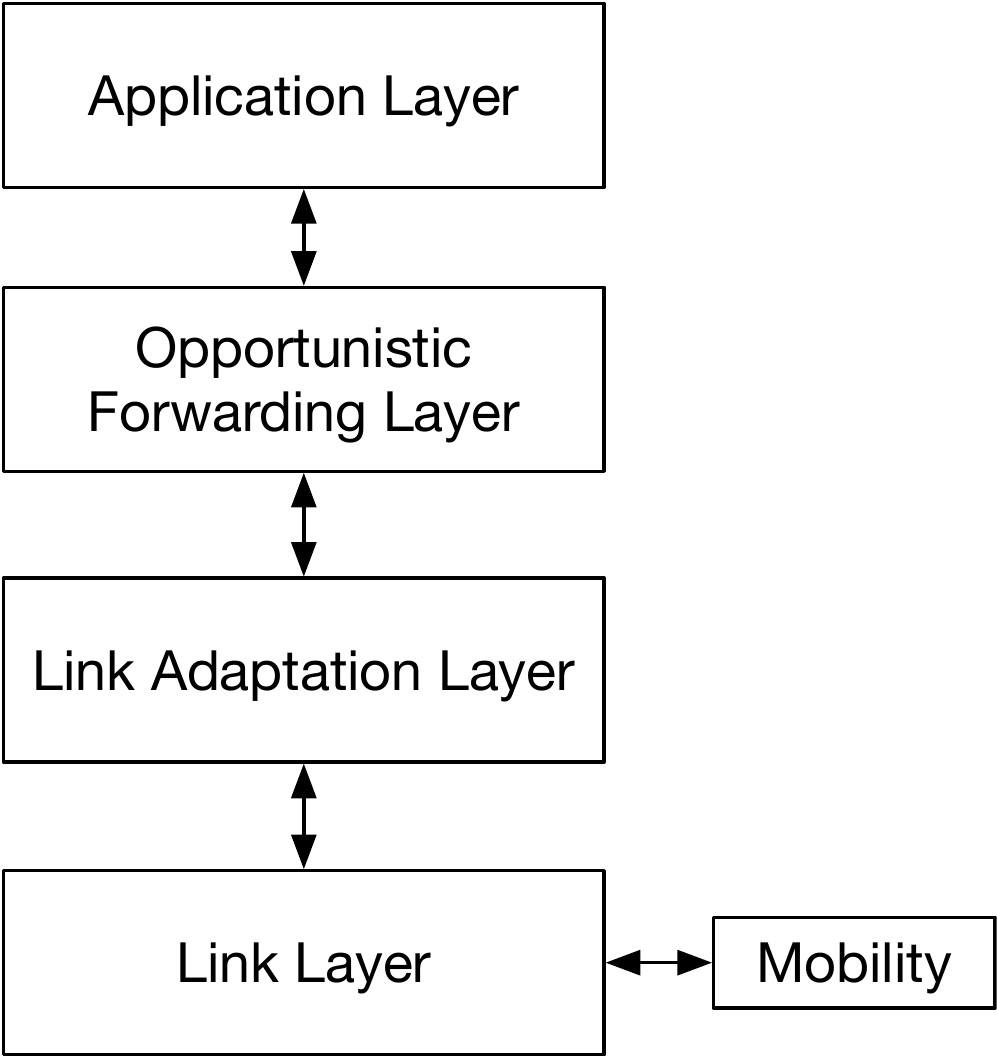}
  \caption{Node Architecture of an OppNets Node}
  \label{fig:proto_stack_block}
\end{figure}

The models of OPS are grouped into protocol layers of a protocol stack. The protocol stack represents the node architecture of a node in an opportunistic network. Each layer focuses on a specific area of operation of an opportunistic networking node. Figure~\ref{fig:proto_stack_block} shows the node architecture of an OppNets node in terms of the protocol stack.  

The \textbf{Application Layer} consist of models that generate data traffic in the network. Currently there are a few applications that were developed to evaluate the performance of OppNets with different data traffic generation patterns. These are listed below.

\begin{itemize}
    \item \textbf{Promote} - This application focusses on generating random data that spread and publicise popular information in a network. A further evaluation done through this application was to understand how different data traffic generation models influence the performance of nodes in the network. There are 3 traffic generation models considered - constant traffic, uniformly distributed traffic and exponentially (poisson) distributed traffic.
    \item \textbf{Herald} - This application, though similar in its operation to the \emph{Promote} application, generates a predetermined set of data in the network, instead of random data. Each data item is assigned a usefulness value by every node before the start of the simulation. The purpose of assigning the usefulness is to determine the percentage of useful data (or liked data) received by every node.
    \item \textbf{Bruit} - This application employs a simple traffic generation model where data traffic is generated using a uniform distribution.
\end{itemize}
 
In all these applications, the actual generation of data (i.e., injecting the data into the network) is assigned to a randomly selected node. A further capability of these applications is to generate destination oriented or destination-less data. In the latter case, the data is meant to simulate information that is liked by multiple users.

The \textbf{Opportunistic Forwarding Layer} is where any forwarding protocol related to disseminating data in opportunistic networks is plugged in. The generic operation of such a protocol is separated into the following parts.

\begin{itemize}
    \item \textbf{Caching of Data} - Unlike in traditional networks, opportunistic networks are characterised by intermittent connections. Therefore, every node must employ a store-and-forward methodology when dealing with data.  
    \item \textbf{Communicating with the Neighbourhood} - In opportunistic networks, the neighbourhood (i.e., nodes in the vicinity with which a given node can communicate) changes constantly. Therefore the forwarding protocol must employ mechanisms to pass-on data to maximize the propagation of this data in the network.
\end{itemize}

The current implementation of OPS includes 3 forwarding protocols. They are listed below.

\begin{itemize}
    \item \textbf{Epidemic Routing} - Epidemic Routing \cite{Vahdat:2002} refers to a data propagation protocol where nodes negotiate with other nodes (in the neighbourhood) to determine unavailable data before the actual data is exchanged. Once a node knows what data it is missing in its cache, these data items are requested one after the other. 
    \item \textbf{Organic Data Dissemination} - The Organic Data Dissemination (ODD) \cite{AF:2015} refers to a technique of  disseminating data based on the popularity of individual data items. It employs a mechanism to determine the significance of change of a node's neighbourhood (i.e., difference between the nodes arrived and left a neighbourhood with respect to a given node) to determine what kind of data is propagated. The \emph{kind of data} refers to the popularity of the data - significant changes result in popular data being propagated.  
    \item \textbf{Randomised Rumor Spreading} - Randomised Rumor Spreading (RRS) is a very simple forwarding protocol to spread data randomly in a network.
\end{itemize}

The \emph{Epidemic Routing} protocol is inherently a \emph{unicast} protocol while the other 2 are broadcast based protocols. The risk of \emph{broadcast} protocols is flooding of networks and these 2 protocols have mechanisms that prevent the flooding of networks.

\begin{figure}[!ht]
  \centering
    \includegraphics[width=0.3\textwidth]{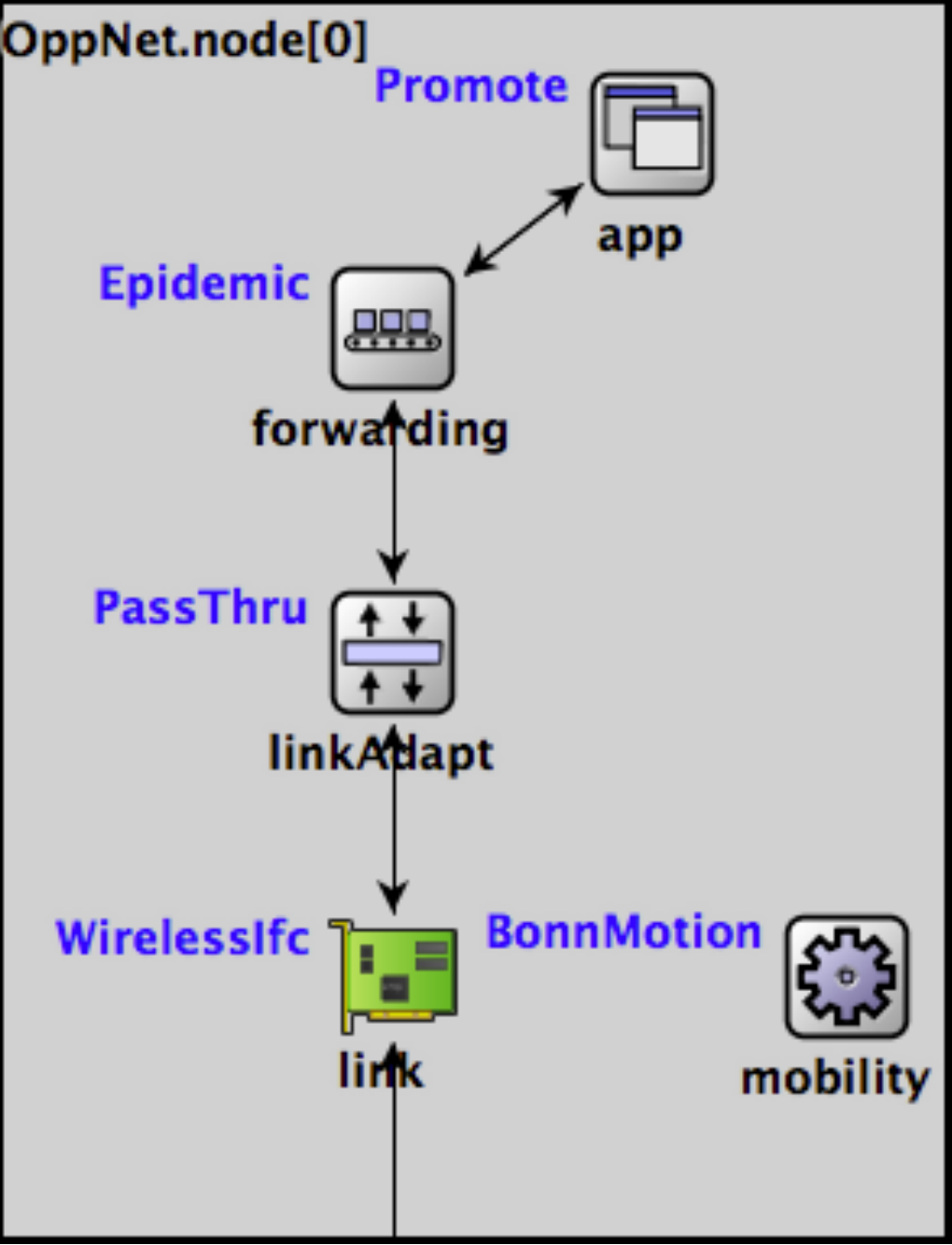}
  \caption{An Example Node Configuration of an OPS Node in OMNet++}
  \label{fig:omnet_ops_node_arch}
\end{figure}

The \textbf{Link Adaptation Layer} is a layer where mechanisms that convert messages from one form to another is employed. An example is tunneling (de-tunnelling) of packes due to the end points of a communication spanning beyond the \emph{Link Layer} end points. Currently, there are no specific adaptation mechanisms implemented. Hence, the \textbf{PassThru} implementation is used at this layer to let packets traverse this layer on its way between the \emph{Opportunistic Forwarding Layer} and the \emph{Link Layer}.

The \textbf{Link Layer} is where any link access technology is implemented. Currently, there is a modeled link layer implementation called \textbf{WirelessIfc}. This implementation, though not a full-fledged wireless link technology, models bandwidth, delays, wireless range and queuing present in a wireless link technology. It uses the Unit Disk Graph (UDG) method as the connectivity model. \textbf{Mobility} of nodes is handled using the mobility interface provided by the \textbf{INET Framework} of OMNeT++. \emph{WirelessIfc} uses the position coordinates of nodes (retrieved from the mobility model) to determine the distance to other nodes. Since it uses the mobility interface of \emph{INET}, it is able to use any mobility model implemented in \emph{INET}.  The \emph{WirelessIfc} has been used to simulate \emph{Bluetooth} network interfaces in our simulations \cite{Dede:2017}.

The interfaces between each layer of the protocol stack has a standard format which has to be adhered to by all protocol implementations.

Figure~\ref{fig:omnet_ops_node_arch} shows an example configuration of the protocol stack of a node in OPS. This configuration uses \emph{Promote} application to generate data traffic. The data dissemination in the opportunistic network is performed using the \emph{Epidemic Routing} protocol. \emph{PassThru} is used to let packets through to other layers. \emph{WirelessIfc} is used to perform wireless peer-to-peer communications and mobility is configured to use the \emph{BonnMotion} mobility model of \emph{INET}. The \emph{BonnMotion} mobility model is a model to move a node using a trace file. A trace file provides the movement pattern in terms of X, Y, and Z coordinates.

OPS is programmed to provide performance data to compute a number of metrics. These metrics are very specific to OppNets to evaluate how well the data is disseminated in a network. The current metrics in OPS are listed below.

\begin{itemize}
    \item \textbf{Liked Data Receipts} - This metric shows the amount of liked data received, compared to all the data received. Liked data are the data that were considered as being useful at the beginning of a simulation. The statistics are computed per node as well as for the whole network.
    \item \textbf{Non-liked Data Receipts} - This metric shows the amount of non-liked data received, compared to all the data received. Non-liked are the data that were not classified as being useful at the beginning of a simulation. The statistics are computed per node as well as for the whole network.
    \item \textbf{Data Traffic Spread} - This metric shows the Coefficient of Variation (CoV) which gives an indication of how well the data traffic is distributed in an opportunistic network.
    \item \textbf{Data Delivery Ratio} - This metric shows the ratio of delivered data to all the data that was generated.
    \item \textbf{Average Delivery Time} - This metric shows how long (at an average) it takes for data to be delivered to the intended recipient. 
    \item \textbf{Average Contact Time} - This metric shows the average time that nodes were in contact during a simulation.
    \item \textbf{Number of Contacts} - This metric shows the times that nodes were in contact during a simulation.
\end{itemize}

As mentioned before, applications in OPS are able to operate as destination oriented or destination less data generators. Therefore, some of these metrics relate to only destination oriented data (e.g., \emph{Data Delivery Ratio}).

%
%


\section{Evaluations}
\label{sec:eval}

OPS has been used in our research to simulate many OppNets based scenarios. \cite{Dede:2017} is a survey of simulators and concepts related to OppNets. This survey used OPS extensively and this section discusses some of the evaluations done with OPS.

\begin{figure}[!ht]
  \centering
    \includegraphics[width=0.48\textwidth]{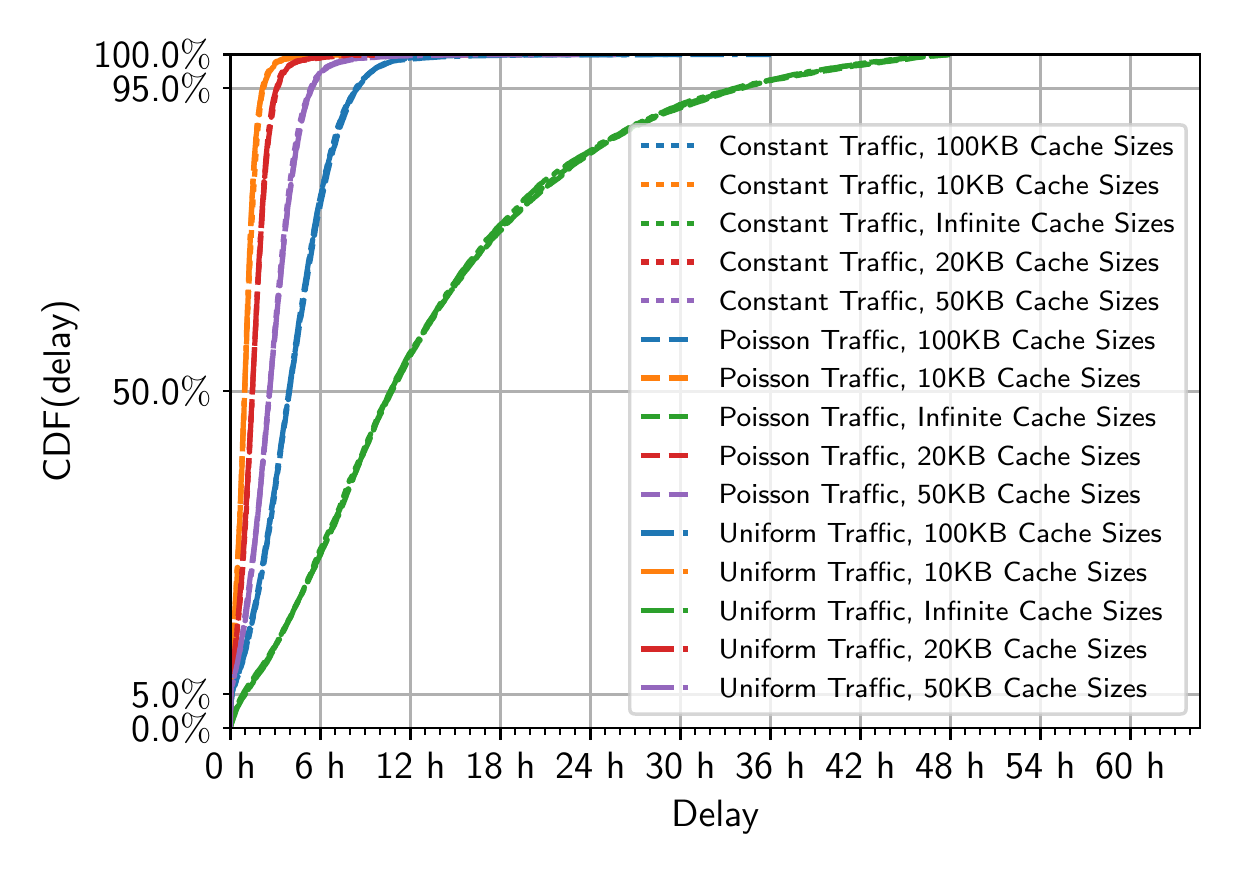}
  \caption{CDF of the Delivery Delay of Data to Recipients using Different Traffic Generation Models and Cache Sizes}
  \label{fig:traffic_experiment_buffersize_1column_omnetpp}
\end{figure}

Figure~\ref{fig:traffic_experiment_buffersize_1column_omnetpp} is the result of simulations done to show the influence of different traffic generation models and cache sizes on the \emph{Average Delivery Time} (referred to as Delay). The results are shown as a set of Cumulative Distribution Function (CDF) curves. The CDFs show that the traffic generation model employed has no influence on the delay. But, the cache size used to store the data influences the \emph{Delay} significantly. 

The simulations for this evaluation is done using a 50-node network. Each of these nodes uses the RRS forwarding protocol of OPS to disseminate data in the network. The mobility of the nodes is performed using the \emph{BonnMotion} mobility model of INET that is configured to use a 50-node extract of the San Francisco taxi cab trace (SFO trace) \cite{Piorkowski:2009}. The simulation time was 30 days and the applications deployed in each of these nodes generate data with a mean interval of 2 hours.

Table~\ref{tab:mobility_results} shows a comparison of the performance of the nodes in an OppNets network when using different mobility models, viz., a simple random model, a trace based model and a hybrid model implemented through Random Waypoint (RWP), Bonn Motion \cite{Aschenbruck:2010} and Small Worlds in Motion (SWIM) \cite{Mei:2009} \cite{udugama:2016}, respectively. The Bonn Motion model uses the SFO trace. As before, the network consist of 50 nodes and a simulation duration of 30 days.

\begin{table}[!ht]
    \begin{center}
        \begin{tabular}{|c||c|c|c|}
            \hline
            \textbf{Model} & \textbf{RWP} & \textbf{SWIM} & \textbf{Bonn Motion} \\
            \hline
            \hline
            \textbf{Simulation Time} & 4 min & 59 min & 109 min \\
            \hline
            \textbf{Memory used} & 74 MB & 86 MB & 127 MB\\
            \hline
            \hline
	        \textbf{Average Delivery Rate} & 3 \%&  96\% & 92 \% \\
            \hline
            \textbf{Average Delivery Delay} & 20.6 h & 16.25 h & 13.16 h\\
            \hline
            \hline
            \textbf{Total Number of Contacts} & 190 & 46,752 & 155,757 \\
            \hline
            \textbf{Average Contact Duration} & 117.14 sec& 150.12 sec & 584.39 sec \\
            \hline
        \end{tabular}
    \end{center}
    \caption{Performance results of different mobility models consisting of a simple random model (RWP), a trace based model (Bonn Motion) and a hybrid model (SWIM), with OPS}
    \label{tab:mobility_results}
\end{table}

The applications deployed on the nodes generate data every 2 hours and the data carry a specific destination to which the data must be delivered. Further, the RWP and the SWIM models are configured as closely as possible to represent the behaviour of the SFO trace. The results show that mobility with real traces require a higer level of resources (simulation clock time and memory usage) compared to other mobility models. When considering the performance of the nodes, it is seen that the SWIM model provides the most closest results (Average Delivery Rate, Average Delivery Delay, etc.) to the trace based model. The trace based model is considered the most realistic due to the traces being collected from real mobility patterns.

The survey \cite{Dede:2017} referenced before included the performance results of other widely used OppNets simulators \cite{Keraenen:2009, Papanikos:2015} in addition to OPS. The results show that OPS provides a comparatively close performance to these simulators when considering the metrics listed in Section~\ref{sec:ops}. 


\section{Conclusion}
\label{sec:conclusion}

The Opportunistic Protocol Simulator (OPS) is an OMNeT++ based modular simulator to evaluate the performance of OppNets. The models in OPS implement the functionality of the different layers of the protocol stack of an OppNets based node. The work presented here provides the details of the different models available in OPS.

OPS is used by us to simulate OppNets based scenarios \cite{Dede:2017} to evaluate the performance of dissemination protocols, traffic generation models, etc., to improve the performance of OppNets in the IoT. OPS is released at Github under a GPL License (\emph{https://github.com/ComNets-Bremen/OPS}).

OPS is being extended with new functionality on a regular basis. Currently, there are a number of on-going projects focusing on areas related to forwarding protocols, applications, mobility models and user behaviour models. A long-term goal is to integrate OPS with the link technologies available with the INET framework.

\bibliographystyle{IEEEtran}
%

\end{document}